# Modeling potent pathways for APC/C inhibition: pivotal roles for MCC and BubR1

Bashar Ibrahim

Bio System Analysis Group, Friedrich-Schiller-University Jena, and Jena Centre for Bioinformatics (JCB),

07743 Jena, Germany

Email: bashar.ibrahim@uni-jena.de

## Abstract

The highly conserved spindle assembly checkpoint (SAC) ensures that the sister chromatids of the duplicated genome are not separated and distributed to the spindle poles before all chromosomes have been properly linked to the microtubules of the mitotic spindle. Biochemically, the SAC delays cell cycle progression by preventing activation of the anaphase-promoting complex (APC/C) or cyclosome; whose activation by Cdc20 is required for sister-chromatid separation, which marks the transition into anaphase. In response to activation of the checkpoint, various species control the activity of both APC/C and Cdc20. However, the underlying regulatory pathways remain largely elusive.

In this study, five possible model variants of APC/C regulation were constructed, namely BubR1, Mad2, MCC, MCF2 and an all-pathways model variant. These models are validated with experimental data from the literature. A wide range of parameter values have been tested to find critical values of the APC binding rate. The results show that all variants are able to capture the wild type behaviour of the APC. However, only one model variant, which included both MCC as well as BubR1 as potent inhibitors of the APC, was able to reproduce both wild type and mutant type behaviour of APC regulation.

The presented work has successfully distinguished between five competing dynamical models of the same biological system using a systems biology approach. Furthermore, the results suggest that systems-level approach is vital for molecular biology and could also be used for compare the pathways of relevance with the objective to generate hypotheses and improve our understanding.

**Introduction**

Correct DNA segregation is a fundamental process that ensures the faithful inheritance of genomic information for the propagation of cell life. Eukaryotic cells have evolved a conserved surveillance control mechanism for DNA segregation called the Spindle Assembly Checkpoint (SAC;(Minshull, et al. 1994)). The SAC is responsible for delaying the onset of anaphase until all chromosomes have made amphitelic tight bipolar attachments to the mitotic spindle.

The SAC works by inhibiting the anaphase-promoting complex (APC/C), presumably through sequestering the ACP/C-activator Cdc20. Upon APC/C activation, a signaling cascade is triggered that is not only responsible for degradation of mitotic cyclins, but additionally causes securin (budding yeast Pds1) to be tagged for degradation by the proteasome. Securin binds and thereby inhibits separase (budding yeast Esp1), a protease required to cleave cohesin, which is the 'glue' connecting the two sister-chromatids of every chromosome. Thus, activation of APC/C by Cdc20 initiates sister-chromatid separation, which marks the transition into anaphase (Musacchio and Salmon 2007). A dysfunction of the SAC can lead to aneuploidy (Suijkerbuijk and Kops 2008) and furthermore its reliable function is important for tumor suppression (Holland and Cleveland 2009,Morais da Silva, et al. 2013).

The central proteins involved in SAC, that are conserved in all eukaryotes, include MAD ("Mitotic Arrest Deficient"; Mad1, Mad2, and Mad3 (in humans: BubR1)) (Li and Murray 1991) and BUB ("Budding Uninhibited by Benzimidazole"; Bub1, and Bub3) (Hoyt, et al. 1991). These proteins work to regulate APC/C activity and its co-activator Cdc20. In addition, the SAC also involves several other proteins that participate in important aspects of this mechanism. Among these additional proteins are Aurora-B (Vagnarelli and Earnshaw 2004) and the "Multipolar spindle-1" protein (Mps1) (Fisk, et al. 2004). These two components are required for SAC signal amplification. Moreover, several other components involved in carrying out essential aspects of the SAC mechanism have been identified in higher eukaryotes, for example, the RZZ complex (Karess 2005,Lu, et al. 2009) which is composed of "Rough Deal" (Rod) (Raff, et al. 2002), Zeste White 10 (Saffery, et al. 2000a,Saffery, et al. 2000b) and Zwint-1(Kops, et al. 2005).

The Cdc20-binding protein Mad2 was suggested as a candidate for the "wait-anaphase" signal, as it is stabilized in a conformation with increased affinity for Cdc20 specifically at unattached kinetochores. The resulting Cdc20:C-Mad2 complex is diffusible and can bind to a complex of Bub3 and BubR1 to form the possibly transient mitotic checkpoint complex (MCC), which is a potent inhibitor of the APC. The MCC inhibits the APC/C in two ways. Firstly, it binds to the APC/C in a way preventing Cdc20 from interacting with mitotic APC-targets (Chao, et al. 2012). Secondly, it directs APC/C-activity towards ubiquitination of Cdc20 (Diaz-Martinez and Yu 2007,Nilsson, et al. 2008).

Cdc20:C-Mad2 can also directly bind to the APC/C and form an inactive complex (Fang, et al. 1998). It has been proposed that Mad2 may act as a catalyst for the assembly of a complex of Bub3:BubR1 with Cdc20 (Kulukian, et al. 2009a,Overlack, et al. 2014). Furthermore, BubR1 was suggested to interact with APC/C (Han, et al. 2013). Another inhibitor, called the mitotic checkpoint factor 2 (MCF2), is associated with APC/C merely in the checkpoint arrested state but its composition is not known (Eytan, et al. 2008). With the exception of MCF2, all complexes inhibiting APC rely on the presence of Cdc20:C-Mad2, which requires unattached kinetochores for adequately fast formation.

Computational models are important tools that help to elucidate how such elaborate systems may work (Ibrahim 2015). So far, mathematical models have helped to enlighten crucial modules of the SAC control mechanism, mainly the "Mad2 template". Doncic et al. (Doncic, et al. 2005) likewise Sear and Howard (Sear and Howard 2006) examined simple spatial models of potential checkpoint mechanisms with focus on budding yeast or metazoan, respectively. They found that a diffusible signal can generally account for checkpoint operation. A more comprehensive model of the human SAC, including the Mad2 and BubR1 pathway has been proposed by Ibrahim et al. (Ibrahim, et al. 2008a,Ibrahim, et al. 2008b,Ibrahim and Henze 2014,Ibrahim, et al. 2009). A mathematical model for checkpoint activation based on the "Mad2 template" model by De Antoni et al. (De Antoni, et al. 2005) is supported with *in vitro* experiments by Simonetta et al. (Simonetta, et al. 2009). Mistry et al. (Mistry, et al. 2008) constructed an elaborated model for the evolution over time of the kinetochore attachment states' distribution. Not only were the attached and unattached states considered, but also the merotelic and syntelic misalignments. Lohel et al. (Lohel, et al. 2009) accomplished a betterment of these models by taking into account species localization and binding sites at kinetochores.

However, no modeling has been considered for the full SAC mechanism including all possible regulatory pathways of the APC. Here, five quantitative model variants for SAC activation and maintenance, considering all known pathways of APC/C regulation in human cells, will be discussed. All core proteins and their reactions are enabled with *in-vitro* measured concentrations and interaction rates. Thus, the simulations give a reliable insight into the behavior of the real SAC system. To mimic the majority of experimental findings, human as well as budding yeast cells are considered. Moreover, various mutations (deletion and overexpression) have been used to distinguish between these model variants.

## Materials and Methods

### Mathematical treatment and simulation

By applying general principles of mass-action kinetics, the reaction rules were converted into sets of time dependent nonlinear ordinary differential equations (ODEs). The actual initial amounts for reaction species are taken from literature (cf. Table 1). When only a certain range was known, different starting values were used for the computations, but qualitatively differing trajectories were not found. The kinetic rate constants are also taken from literature as far as they are known. In the other cases, representative values exemplifying their whole physiologically possible range were selected. A summary of all simulation parameters is given in Table 1.

In a typical simulation run, all reaction partners were initialized according to Table 1 and numerically integrated the ODEs until steady state was reached before attachment (using u = 1, see chemical reaction scheme). After switching u to 0, the equations are again integrated, until steady state is reached. To solve the non-linear ODEs systems, a variety of computational approaches have been applied. In particular, high order (seven–eighth) Runge–Kutta (Fehlberg and Prince) formulas, a modified Rosenbrock formula, variable order Adams–Bashforth–Moulton, backward differentiation formulas, and an implicit Runge–Kutta method. Switch between these methods are done automatically according to the system's stiffness state. This helps to simulate wide range of parameters (e.g. $10^3$-$10^9$) and also validate our solutions. The implementation and simulations code are written based on MATLAB (Mathworks, Natick, MA).

**Model assumptions**

Some reactions can depend on the attachment status of the kinetochores, so all reactions can be classified by whether they upon microtubule attachment are unaffected ("uncontrolled"), turned off ("off-controlled") or turned on ("on-controlled"). Only reactions involving kinetochore localized species can be controlled. For example, formation of Mad1:C-Mad2:O-Mad2* (Reaction 2) can only take place as long as the kinetochores are unattached. In this model, if the kinetochore is unattached, u was set to u = 1, otherwise u = 0. Note that it is assumed that mass-action-kinetics is used for all reactions (Ibrahim, et al. 2008a,Ibrahim, et al. 2008b). If other kinetics should be used, they have to be transformed to mass action kinetics.

Mad1:Mad2 is considered to be preformed complex and the complex formation is not considered. It should be noted that this complex is a tetrameric 2:2 Mad1:Mad2 and not a monomer complex. From mathematical point of view, considering the complex as a species would not make any difference in this case as long as there is one form in our model. All previous mathematical models have considered the similar assumption to the template model (e.g.,(Ibrahim, et al. 2008a,Lohel, et al. 2009,Simonetta, et al. 2009)).

**Results and Discussion**

**Biochemical background of the model**

The reaction network of the SAC activation and maintenance mechanism can be divided into kinetochore dependent and cytosolic (kinetochore independent) parts. The former serves to communicate the attachment status of each kinetochore to the remainder of the cell while the latter accounts for the actual inhibition of the APC.

**Template model of Mad2-activation**

The protein Mad2 is present in two stable conformations differing in the spatial arrangement of a 'safety-belt' which is either open (O-Mad2) or closed (C-Mad2) (De Antoni, et al. 2005,Luo, et al. 2004,Sironi, et al. 2002). O-Mad2 is able to bind Cdc20, though the resulting complex is rather transient. C-Mad2 cannot bind to Cdc20, but in contrast, the complex of C-Mad2 and Cdc20 is quite stable. Central to the SAC-network is a kinetochore-bound template complex made up from Mad1 and C-Mad2. This template complex recruits O-Mad2 and stabilizes an intermediate conformation (O-Mad2*) which can bind Cdc20 efficiently and switches to closed conformation upon Cdc20-binding, tightening the connection between both with the 'safety-belt' (De Antoni, et al. 2005,Luo, et al. 2004,Vink, et al. 2006). The C-Mad2:Cdc20 complex formed by this mechanism has been given the name "template-model" (De Antoni, et al. 2005).

Mad2-activation at the kinetochores is commonly seen as the central element of the SAC. Therefore the template model of Mad2-activation is included into the model (De Antoni, et al.

2005,Ibrahim, et al. 2008b,Simonetta, et al. 2009). The kinetic rate coefficients for this interactions have been determined *in vitro* by Simonetta et al. (2009). The biochemical equations of the Mad2 Template Model are described by reaction (R1)–(R3) (see chemical reaction scheme, below).

**Autocatalytic amplification of Cdc20:C-Mad2 formation**

In addition to its activation via kinetochore-bound Mad1:C-Mad2, O-Mad2 can likewise be activated by Cdc20:C-Mad2 to autocatalytically increase Cdc20:C-Mad2 formation rate *in vitro* (Simonetta, et al. 2009). However, the contribution of this autocatalytic loop is minor with the kinetic data given in (Simonetta, et al. 2009). The presence of a highly contributing autocatalytic loop would also counteract checkpoint deactivation and is therefore not desirable for live cells from a theoretical point of view (Ibrahim, et al. 2008b). This loop is therefore not considered in the model. For integrity, the biochemical equations of the autocatalytic amplification are listed by reaction (R4)–(R5) (see chemical reaction scheme, below).

**MCC formation**

Although Cdc20:C-Mad2 can bind to and inhibit APC directly, its inhibitory potency increases greatly in synergy with the Bub3:BubR1 complex (Musacchio and Salmon 2007). Cdc20:C-Mad2 together with Bub3:BubR1 was shown to form the tetrameric mitotic checkpoint complex (MCC), which is a potent inhibitor of APC (Sudakin, et al. 2001). The trimeric complex Bub3:BubR1:Cdc20 is a potent inhibitor of APC, too (Malureanu, et al. 2009). However, the rate of its uncatalyzed formation in the cytosol is slow (Sudakin, et al. 2001). The formation of Bub3:BubR1:Cdc20 is accelerated in the presence of unattached chromosomes (Kulukian, et al. 2009b) and it may be that MCC forms as an intermediate complex from which O-Mad2 rapidly dissociates (Kulukian, et al. 2009b,Malureanu, et al. 2009,Medema 2009). Microtubule attachment depletes the template-complexes from the respective kinetochore, thereby silencing

SAC-signaling locally (Buffin, et al. 2005,Sivaram, et al. 2009). Thus, after proper attachment of the last chromosome, SAC signaling ceases and passage to anaphase is granted. The MCC formation is described by the reaction (R6) (see chemical reaction scheme, below).

**APC inhibition**

The APC can be inhibited in multiple ways, and complexes of APC together with either Cdc20:C-Mad2, Bub3:BubR1, Bub3:BubR1:Cdc20, MCC or MCF2 have been found to be inactive (Eytan, et al. 2008,Herzog, et al. 2009,Kulukian, et al. 2009b,Malureanu, et al. 2009,Medema 2009,Sudakin, et al. 2001). However, the demand mechanisms for binding the inhibitory complexes to the APC are subject to current research.

The MCC may form more stably at unattached kinetochores and recruit APC from the cytoplasm (Herzog, et al. 2009). The MCC sub-complexes Bub3:BubR1:Cdc20 and Cdc20:C-Mad2 can bind to the APC independently from unattached kinetochores although their binding may be facilitated in a kinetochore-dependent manner (Kulukian, et al. 2009b,Malureanu, et al. 2009,Medema 2009). The MCC may also bind to APC in a kinetochore independent manner to eventually inhibit APC by releasing O-Mad2, forming the stable Bub3:BubR1:Cdc20:APC complex. The recently discovered mitotic checkpoint factor 2 protein (MCF2) is a highly potent APC-inhibitor, yet the mechanism of binding to the APC and its regulation is still elusive (Braunstein, et al. 2007,Eytan, et al. 2008). All complexes inhibiting APC (except of MCF2) rely on the presence of Cdc20:C-Mad2, which requires unattached kinetochores for adequately fast formation. The Cdc20:C-Mad2 complex can be characterized to be the "interface" connecting signaling from unattached kinetochores to APC inhibition. The APC regulation reactions are described by the reactions (R8)–(R10) (see chemical reaction scheme, below).

**Chemical reaction scheme**

In this model of the SAC mechanism, 13 biochemical reaction equations describe the dynamics of the following 16 species: Mad1:C-Mad2, O-Mad2, Mad1:C-Mad2:O-Mad2*, Cdc20, Cdc20:C-Mad2, Bub3:BubR1, MCC, Bub3:BubR1:Cdc20, APC, MCC:APC, APC:BubR1:Bub3, APC:Cdc20:C-Mad2, APC: Cdc20:BubR1:Bub3, MCF2, MCF2:APC, and APC:Cdc20.

$$\text{Cdc20} + \text{O-Mad2} \underset{k_{-1}}{\overset{k_1}{\rightleftarrows}} \text{Cdc20:C-Mad2} \tag{R1}$$

$$\text{Mad1:C-Mad2} + \text{O-Mad2} \underset{k_{-2}}{\overset{k_2 \cdot u}{\rightleftarrows}} \text{Mad1:C-Mad2:Mad2*} \tag{R2}$$

$$\text{Mad1:C-Mad2:Mad2*} + \text{Cdc20} \underset{k_{-3}}{\overset{k_3 \cdot u}{\rightleftarrows}} \text{Mad1:C-Mad2} + \text{Cdc20:C-Mad2} \tag{R3}$$

$$\text{Cdc20:C-Mad2} + \text{O-Mad2} \underset{k_{-4}}{\overset{k_4}{\rightleftarrows}} \text{Mad1:C-Mad2:Mad2*} \tag{R4}$$

$$\text{Cdc20:C-Mad2:Mad2*} + \text{Cdc20} \underset{k_{-5}}{\overset{k_5}{\rightleftarrows}} 2\text{Cdc20:C-Mad2} \tag{R5}$$

$$\text{Cdc20:C-Mad2} + \text{BubR1:Bub3} \underset{k_{-6}}{\overset{k_6 \cdot u}{\rightleftarrows}} \text{MCC} \tag{R6}$$

$$\text{Cdc20} + \text{BubR1:Bub3} \underset{k_{-7}}{\overset{k_7}{\rightleftarrows}} \text{Cdc20:BubR1:Bub3} \tag{R7}$$

$$\text{MCC} + \text{APC} \underset{k_{-8}}{\overset{k_8 \cdot u}{\rightleftarrows}} \text{MCC:APC} \tag{R8}$$

$$\text{APC} + \text{Cdc20} \underset{k_{-9}}{\overset{k_9}{\rightleftarrows}} \text{APC:Cdc20} \tag{R9}$$

$$\text{APC} + \text{Cdc20:C-Mad2} \underset{k_{-10}}{\overset{k_{10} \cdot u}{\rightleftarrows}} \text{APC:Cdc20:C-Mad2} \tag{R10}$$

$$\text{APC} + \text{BubR1:Bub3} \underset{k_{-11}}{\overset{k_{11} \cdot u}{\rightleftarrows}} \text{APC:BubR1:Bub3} \tag{R11}$$

$$\text{APC} + \text{Cdc20:BubR1:Bub3} \underset{k_{-12}}{\overset{k_{12} \cdot u}{\rightleftarrows}} \text{APC:Cdc20:BubR1:Bub3} \tag{R12}$$

$$\text{MCF2} + \text{APC} \underset{k_{-13}}{\overset{k_{13} \cdot u}{\rightleftarrows}} \text{MCF2:APC} \tag{R13}$$

**Dynamics of APC/C regulation**

Five model variants for the regulation of APC/C at meta- to anaphase transition have been constructed. As described in the literature, many proteins contribute to checkpoint function. The key players and their interactions are captured by the reaction equations introduced in the previous section. These equations are then transformed into ODEs and selected specific values for the initial concentrations and rate constants from the literature and our previous publications (summarized by Table 2). All model variants share the core template model of Mad2-activation (reactions (1)-(3)), MCC formation (Reactions (6)-(7)), MCC binding to APC (Reaction (8)), and APC activation by Cdc20 co-activator (Reaction 9) (see text above). They differ by the additional inhibitory pathway for the APC. Hence, the basic model has solely MCC as a direct binding partner and inhibitor to APC, namely the MCC dominated model variant (Reactions (1)-(9)). The second model variant consists of the basic MCC model as well as Mad2 as a direct inhibitor of APC, namely Mad2 dominated model variant (Reactions (1)-(10)). The third model variant the basic model and also BubR1 as direct inhibitor, the BubR1 dominated pathway. This pathway has two variants either binds APC (Reactions (1)-(9), and (11)) or BubR1:Cdc20 complex which is binds to APC (Reactions (1)-(9), and (12)). The fourth model contains MCF2 inhibition of APC in addition to MCC basic model (Reactions (1)-(9), and (13)). The fifth model variant includes all of the listed above variants simultaneously.

In all simulations, APC:Cdc20 concentration is considered as comparison reference which should be close to zero before attachment and should rise quickly after attachment (spindle attachment occurs at t = 2000s). Reactions that are depending on kinetochore signal are switched from 1 to zero after attachment. All results are presented for different values of the rate $k_8$ (MCC binding to APC). Parameters setting are according to Table 1. Simulation shows that all other pathways in this study should be controlled same as MCC pathway. The type of control is unknown and kinetochore should be directly or indirectly involved. Thus, only the controlled pathway types will be shown in the simulation figure.

### MCC dominated model (MCC-APC pathway)

The SAC model has been previously analyzed where MCC is the exclusive inhibitor of APC (Ibrahim, et al. 2008a). This model is used in this study as a core part (Reactions (1)-(9)) to build on and compare with other model variants. First, this basic model was re-simulated with various MCC-APC binding rates ($10^3$ to $10^9$ $M^{-1}s^{-1}$). For all calculations, the concentrations and rates of Table 1 were chosen. In Figure 1, the APC:Cdc20 (active APC form) concentration is plotted for seven different values of $k_8$ (MCC-APC binding rate). With increasing values of $k_8$, a faster switching behavior is observed. The MCC shows sufficient inhibition of APC only when the binding rate is high enough (Fig. 1A). This could be the case; however, there is neither experimental data nor a theoretical indication showing the MCC-APC binding rate, yet.

### Mad2 dominated model (Mad2:Cdc20-APC pathway)

In this model variant, in addition to the MCC model pathway (Reactions (R1)-(R9)), Mad2-Cdc20 complex direct binding to APC (Reaction R10) is added. When simulating this model, considering the Reaction 10 binding rate the same as the MCC binding rate (Reaction 8), no changes were observed (same as Fig. 1A). As the Mad2 binding was increased gradually, it was discovered that the binding rate needed to be 1000 fold higher than MCC binding to APC in order to reach ideal inhibition (zero before attachment and max after attachment). These results are shown in Fig. 1B.

### BubR1 dominated model (BubR1- or BubR1:Cdc20-APC pathways)

This model variant has two possibilities; the first variant is one where BubR1 directly binds the APC (Reaction R11) and second is where the BubR1:Cdc20 complex binds the APC (Reaction R12). The simulation of reaction (R1-R11) results in no effect as long as the binding rate is similar to MCC binding to APC. Once the rate increased 100 fold than the MCC-APC binding rate, the inhibition of APC reached about zero level (Fig.1C). This value is exactly 10 fold less than the required rate for Mad2 binding to APC. This may imply that BubR1 is 10 fold more potent than Mad2 in inhibiting APC. Interestingly, this value is very similar to the experimental finding (Fang 2002) where they found that BubR1 is 12 fold more potent than Mad2. However, this could be incidental.

The second variant where the BubR1:Cdc20 complex binds the APC, (R1-R10, and R12) shows no improvement for APC inhibition or recovery even with a very high binding rate ($10^5 M^{-1}s^{-1}$, Fig.1D, and compare to Fig.1A). This may imply that BubR1:Cdc20 is less likely to act as a direct inhibitor of APC while BubR1 alone does.

**MCF2 regulation model (MCF2-APC pathway)**

This variant considers MCF2 as a second direct binding partner to the APC in addition to the MCC (R13). The amount of MCF2 is still unknown and so the simulation was run twice; once when its amount is as half the amount of the APC and second where its amount is exactly as the APC amount. Additionally, the binding rate is like the MCC-APC binding rate.

The simulations are summarized in Fig. 1E and Fig. 1F. To achieve full inhibition of the APC, a high amount (same as APC amount which is 0.09 µM) as well as a high binding rate ($10^5 M^{-1}s^{-1}$) is required.

**All pathways simultaneously (MCC, Mad2, BubR1, and MCF2)**

There is still the possibility where all pathways regulate the APC simultaneously (Reaction R1-R13). The simulation results of this model show ideal inhibition of the APC (very low before attachment and dramatically high after) with the condition that the binding rates are about $10^6$.

Taken together, to this end, any additional pathway can increase the inhibition of APC. All the variant models show the desired SAC mechanism. However, further analysis must be carried out to discriminate between these model variants.

**Model validation by mutation experiments**

In order to validate all model variants and find out their limitation and constraints, different mutations (deletion and over-expression) of the proteins involved (Table 2) were tested. There are many experimental studies reported where deletion and also overexpression in different organisms of any of the core components, Mad2 (De Antoni, et al. 2005,Dobles, et al. 2000,Fang, et al. 1998,He, et al. 1997,Kabeche and Compton 2012,Michel, et al. 2001,Nath, et al. 2012,Nezi, et al. 2006), BubR1 (Chan, et al. 1999,Davenport, et al. 2006,Harris, et al. 2005,Ouyang, et al. 2002,Yamamoto, et al. 2007), and Cdc20 (Hwang, et al. 1998,Mondal, et al. 2006,Mondal, et al. 2007,Shirayama, et al. 1999,Zhang and Lees 2001), resulted in SAC defects, such as failed or successful mitotic arrest. These experiments helped in validating all model variants and additionally discriminating between them. The experiments from literature are listed in Table 2.

In the simulations, the respective initial concentration was set to zero for the deletions, and 100 fold higher concentrations for over-expression. The desired proper wild type functioning, APC:Cdc20 concentration should be very low (zero) before the attachment, and should increase quickly after attachment. Cells failing to arrest meant a very high level of APC:Cdc20 and low sequestration level of Cdc20. Arrested cells meant a very low level of APC:Cdc20 and full sequestration of Cdc20. The simulations show that all the above model variants are unable to fully reproduce all known experimental findings for the given parameter set (Table 2). Thus, the simulations were re-run with a larger range of parameter values and these results are summarized in Fig. 2.

The MCC dominated model variant is shown in the first row of Fig. 2. Conjointly, deletion of Mad2 or BubR1 resulted in high APC:Cdc20 levels (0.09 µM) for any parameter set ($10^3$-$10^9$ $M^{-1}s^{-1}$) while overexpression of Cdc20 required that the APC binding rate must be less than $10^6$ $M^{-1}s^{-1}$.Overexpression of Mad2 or deleting Cdc20 resulted in low levels of APC:Cdc20 (very near to zero) for any parameter values while overexpression of BubR1 required that the APC binding rate to be higher than $10^6$ $M^{-1}s^{-1}$. Taken together, for the MCC model variant, the APC-MCC binding rate must be exactly $10^6$ $M^{-1}s^{-1}$. This rate is 10 fold less than the theoretical reported prediction (Ibrahim, et al. 2008a). With this rate, the wild type simulation cannot be met (low APC:Cdc20 before attachment and high after), see Fig. 1.A. Thus this model variant could be excluded.

The simulation in the second row (Fig. 2) shows the BubR1 dominated variants (either BubR1-APC or Cdc20:BubR1-APC). Both model variants acted rather similarly for BubR1, Cdc20 deletion and Cdc20 or Mad2 overexpression. They differed in the deletion of Mad2 or overexpression of Cdc20. The BubR1-APC variant requires that the binding rate to be less than $10^4$ $M^{-1}s^{-1}$ whereas the Cdc20:BubR1-APC variant requires that the rate to be less than $10^6$ $M^{-1}s^{-1}$. The later variant contradicted the requirements for the wild type simulation (see Fig. 1D). Therefore, Cdc20:BubR1-APC can be excluded.

The Mad2 model variant is shown in the third row Fig. 2. There was no parameter set that can be used to reproduce all mutation experiments. For instance, BubR1 deletion needs a very low binding rate ($10^4$ $M^{-1}s^{-1}$) while overexpression of BubR1 requires high rate (higher than $10^4$ $M^{-1}s^{-1}$). Hence, Mad2 model variant is less likely to be a realistic pathway and may be excluded.

The MCF2 variant (Fig. 2, fourth row) has no parameter value allowing it to reproduce the experimental results, so this model can be excluded. Also, a model where all variants are included was tested (last row, Fig. 2). This model indeed shows the fact that all pathways cannot work in parallel. For example, deletion of BubR1 requires a binding rate of less than $10^4$ $M^{-1}s^{-1}$ while BubR1 overexpression requires binding rate of larger than $10^4$ $M^{-1}s^{-1}$. The wild type simulation for all model variants with low and high binding parameter values are shown in Figure 3. Taken together, only the BubR1-APC model provided an ideal SAC functioning and was able to reproduce all experimental findings. This model has the MCC and BubR1 as a direct inhibitor of APC.

**Conclusion**

The SAC is an intriguingly sensitive checkpoint. The SAC works by inhibiting the APC:Cdc20 complex, which triggers a signaling cascade resulting in sister-chromatid separation upon activation. The SAC delays mitotic progression even if only one kinetochore is not properly attached to the spindle (Musacchio and Salmon 2007). Though, it is still puzzling as to which molecular mechanisms can achieve reliable mitotic arrest while still remaining highly responsive. To complicate the matter even further, the SAC must link the biomechanics of the mitotic spindle with a biochemical signal transduction network, thus posing an inherently spatial problem.

Quantitative analysis and computational modeling are very important tools to elucidate how such elaborate systems work (Ibrahim 2015). So far, mathematical models have helped to elucidate the kinetochore structure and with that the mitotic checkpoint mechanism (Doncic, et al. 2005,Görlich, et al. 2014,Ibrahim, et al. 2007,Ibrahim, et al. 2008b,Ibrahim and Henze 2014,Ibrahim, et al. 2013,Kreyssig, et al. 2012,Kreyssig, et al. 2014,Lohel, et al. 2009,Sear and Howard 2006,Simonetta, et al. 2009,Tschernyschkow, et al. 2013). However, none of these models consider the APC regulatory pathways.

Which component is able to bind and inhibit APC, is a crucial question for SAC regulation. It is a general problem of simulation studies to identify all reaction schemes and to incorporate them with all the correct reaction kinetics. In general, the method of analysis described here can help to falsify false assumptions, but can hardly prove the correctness of a pathway (Popper 1935). In this analysis, it is found that the APC inhibition by only MCC or MCC in addition to MCF2 or Mad2 is very unlikely to describe SAC regulation. However, the direct inhibition of the APC by both BubR1 and MCC could be a basis for further pathway design. The influence of Cdc20 sequestration on SAC operation has been previously analyzed (Ibrahim, et al. 2008b,Musacchio and Salmon 2007). How far partial inhibition of the APC/C by the SAC is tolerable by mitotic control is currently unclear. As many of the basic reactions described here and previously (Ibrahim, et al. 2008a) and for the experimentally determined values, do not lead to complete Cdc20 sequestration, the studies described here have suggested that the MCC directly binds and completely blocks the APC/C. APC/C activation can then be a consequence of a MCC:APC/C complex modification or rearrangement (Ibrahim, et al. 2008a). To investigate and decide on the further reaction scheme for direct inhibition of the APC, additional direct inhibitors were integrated into various models describing APC/C inhibition after metaphase to anaphase transition, this may serve as a basis for more complex investigations. An example in this context is the discovery of a pharmacologic, competitive inhibitor of both Cdc20 and Cdh1 tosyl-L-arginine methyl ester (TAME, (Zeng, et al. 2010)). Also, a cell permeable pro-drug, pro-drug, proTAME, which causes a SAC-dependent mitotic arrest via perturbs the function of the APC/C (Zeng, et al. 2010). These data can be integrated in the presented MCC-BubR1 model. However, to make detailed analysis and perditions, more kinetic data is required such as binding rate for APC/C:Cdc20 in cells treated with proTAME/ or TAME, and other doses of the drug compared to the level of APC/C amount (c.f., 0.1µM - 1µM). Nevertheless, to have an idea, the TAME-model, APC:Cdc20 interaction is disrupted, was integrated into MCC-BubR1 model. The simulation results supported the view that this variant is SAC-dependent under the condition that TAME inhibition reduces at least one fold the APC:Cdc20 interaction rate ($10^4$ $M^{-1}s^{-1}$). Under this condition, mutation experiments are also reproducible (data not shown).

In order to accelerate the pace of molecular biology knowledge, rigorous theoretical analysis should be developed to link computational models of biological networks to experimental data in tight rounds of analysis and synthesis in an integrative systems biology framework (Ibrahim 2015). It is anticipated that a systems biological approach of the SAC mechanism will serve as a basis to integrate future findings and evaluate novel hypotheses related to checkpoint architectures and regulation.

**Acknowledgment**

I acknowledge Fouzia Ahmad for critically reading the manuscript. This work was supported by the European Commission HIERATIC Grant 062098/14.

**Author Disclosure Statement**

The author declares that no conflicting financial interests exist.

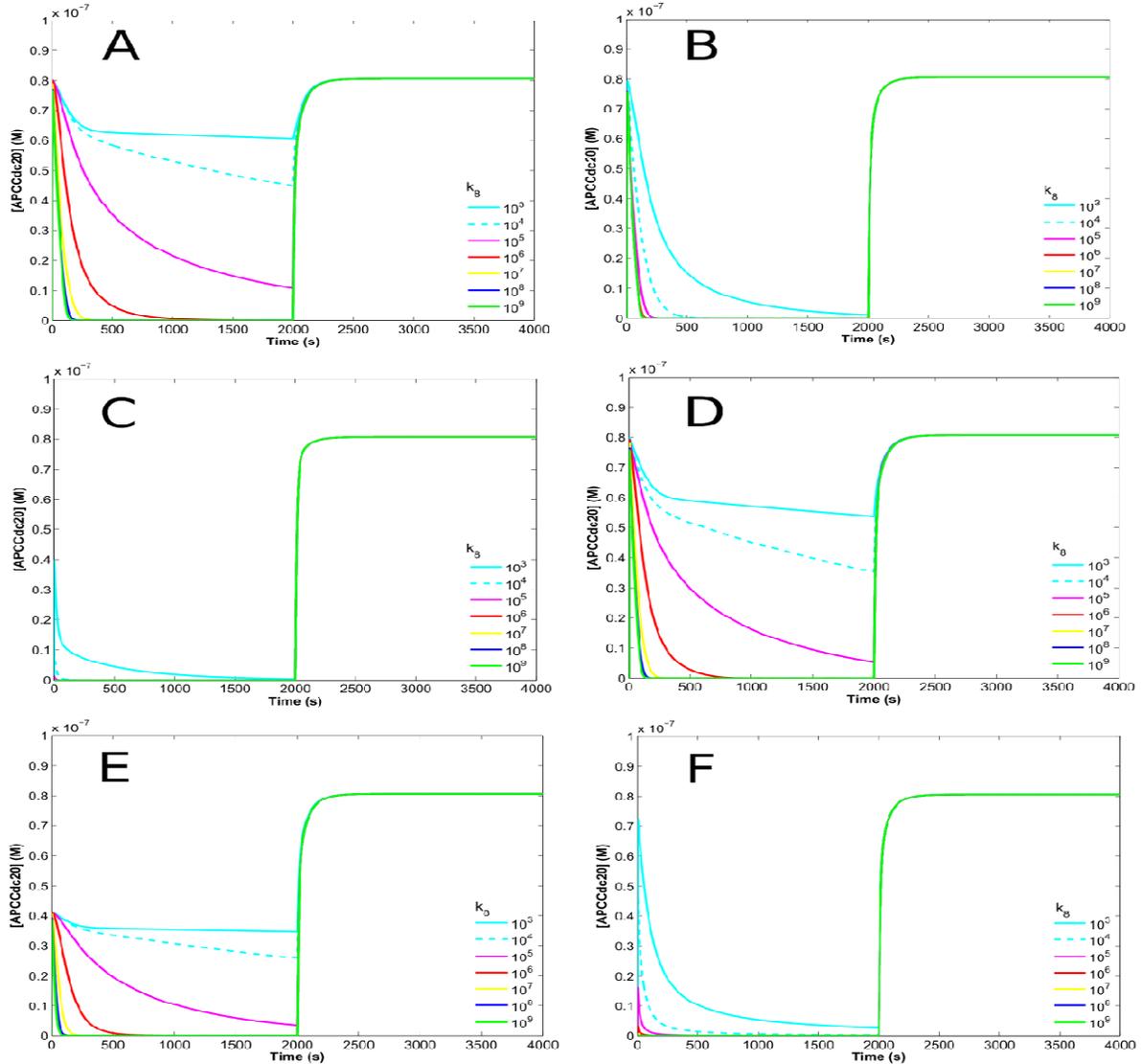

**Figure 1: Dynamical behavior of APC:Cdc20 concentration versus time.** The APC:Cdc20 concentration should be close to zero before attachment and should rise quickly after attachment (spindle attachment occurs at t = 2000s). All results are presented for different values of the rate $k_8$ (MCC binding to APC). Parameters setting are according to Table 1. All other variant show fast recovery and differ in the impact of inhibition.

The MCC basic model variant, **Panel A**, shows fast recovery and only with high MCC binding rate to APC shows fast inhibition. **Panel B** depicts the simulation of Mad2 dominated model variant (Reaction 10). It shows fast inhibition when Mad2:Cdc20 binding rate to APC is $10^4$folds higher than the MCC binding to APC. **Panel C** contains the simulations of BubR1dominated model variant (Reaction 11) which also shown inhibition to APC with rate $10^3$folds higher than the MCC binding to APC. **Panel D** is similar to panel C and the same model variant where BubR1:Cdc20 directly binds APC (Reaction 12). In the later case, no better inhibition of APC can be archive. The simulation for MCF2 dominated model variant (Reaction 12) is shown in **Panel E** and **F**. The concentration of MCF2 is 50% of APC concentration in panel E while the concentration is same as APC concentration in panel F.

## MCC dominated
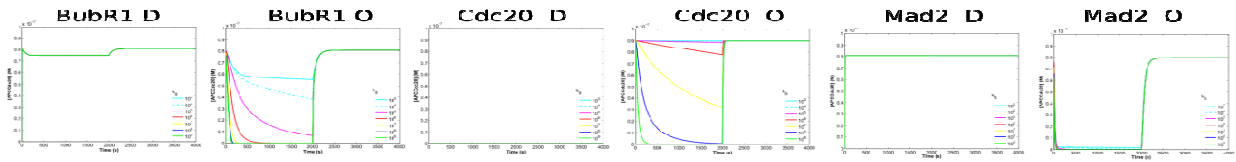

## BubR1 dominated
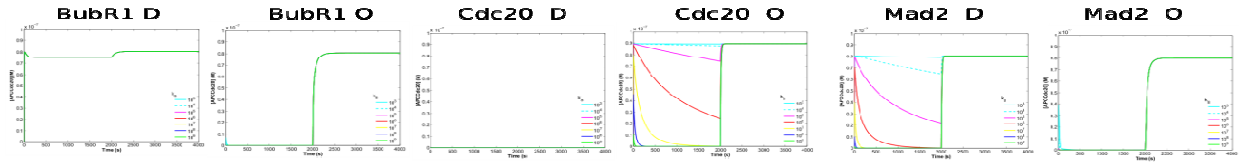

## BubR1:Cdc20 dominated
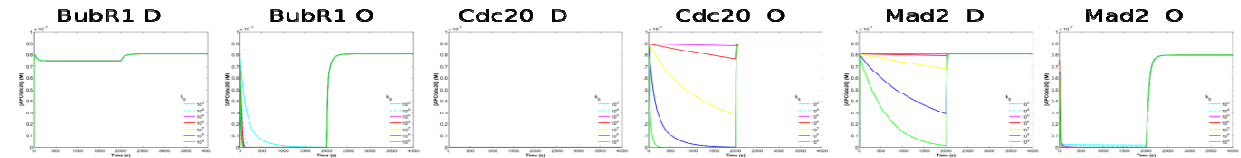

## Mad2 dominated
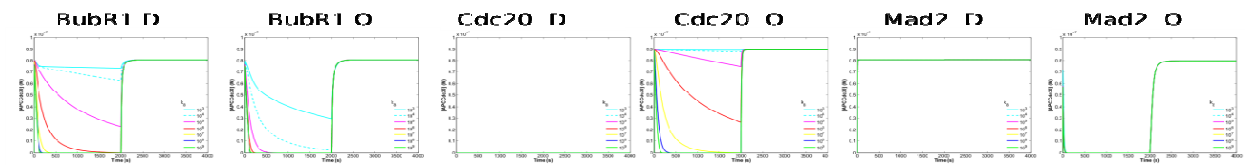

## MCF2 dominated
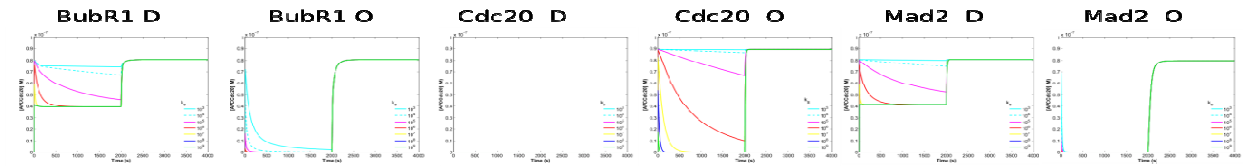

## All variants simultaneously
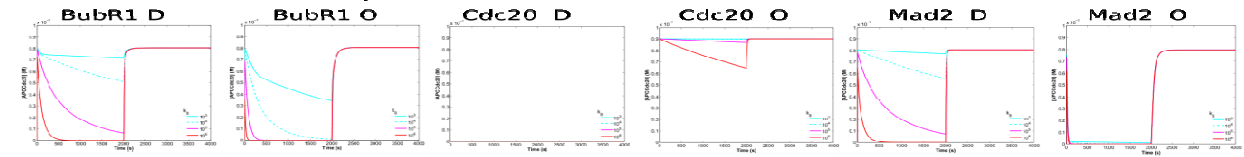

**Figure 2: Simulation of Mad2, BubR1, and Cdc20 mutations for each model variant.**
The respective initial concentration was set to zero for deletion, and 100folds higher for over-expression. Towards proper wild type functioning, APC:Cdc20 concentration should be very low (zero) before the attachment, and should increase quickly after attachment. Deletion of Mad2 or BubR1 or an overexpression of Cdc20 leads to inability of the cell to arrest, that is in the simulation, the concentration of APC:Cdc20 keeps high. Overexpression of Mad2 or BubR1 or deleting Cdc20 results in arresting the cell, that is, the concentration of APC:Cdc20 is very low or zero.
Each row represents the mutation simulations of a model variant and a range of parameter rate for APC binding. Spindle attachment occurs at t = 2000s (switching parameter u from 1 to 0, see R2 for example). All parameters setting are according to Table 1. Each color corresponds to parameter value as in the legend (See text for more details).

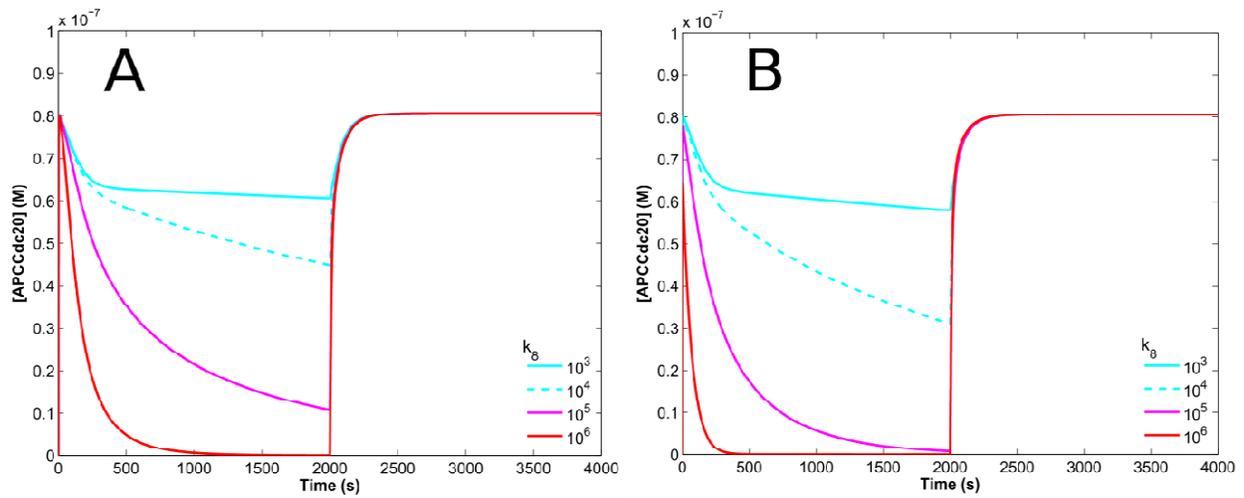

**Figure 3**: **Simulation of APC:Cdc20 after model refinement**. The APC:Cdc20 concentration should be close to zero before attachment and should rise quickly after attachment (spindle attachment occurs at t = 2000s). All results are presented for different values of the rate $k_8$ (MCC binding to APC). Parameters setting are according to Table 1.
**Panel A** shows typical simulation for any of the model variant alone with low binding rates (validated from mutation experiments). **Panel B** shows the simulation where all pathways were included in one model. In the later, MCF2 has no effect whether its level high or low.

**Table 1: Model parameters**

| | *Parameters* | | *Remarks* |
|---|---|---|---|
| **Rate constants** | | | |
| | $k_1$ | $1 \times 10^3$ $M^{-1}s^{-1}$ | (Ibrahim, et al. 2008b, Musacchio and Salmon 2007) |
| | $k_2$ | $2 \times 10^5$ $M^{-1}s^{-1}$ | (Howell, et al. 2000, Vink, et al. 2006) |
| | $k_3$ | $1 \times 10^7$ $M^{-1}s^{-1}$ | (Ibrahim, et al. 2008b) |
| | $k_4$ | $2 \times 10^4$ $M^{-1}s^{-1}$ | (Ibrahim, et al. 2008b, Ibrahim and Henze 2014) |
| | $k_5$ | $1 \times 10^7$ $M^{-1}s^{-1}$ | (Ibrahim, et al. 2008b, Ibrahim and Henze 2014) |
| | $k_6$ | $10^3$-$10^9$ $M^{-1}s^{-1}$ | This study |
| | $k_7$ | $1 \times 10^4$ $M^{-1}s^{-1}$ | (Ibrahim, et al. 2008a, Ibrahim, et al. 2009) |
| | $k_8$ | $1 \times 10^8$ $M^{-1}s^{-1}$ | (Ibrahim, et al. 2008a) |
| | $k_9$ | $5 \times 10^6$ $M^{-1}s^{-1}$ | (Ibrahim, et al. 2008a) |
| | $k_{10}$ | $10^3$-$10^9$ $M^{-1}s^{-1}$ | This study |
| | $k_{11}$ | $10^3$-$10^9$ $M^{-1}s^{-1}$ | This study |
| | $k_{12}$ | $10^3$-$10^9$ $M^{-1}s^{-1}$ | This study |
| | $k_{13}$ | $10^3$-$10^9$ $M^{-1}s^{-1}$ | This study |
| | $k_{-1}$ | $1 \times 10^{-2}$ $s^{-1}$ | (Ibrahim, et al. 2008b) |
| | $k_{-2}$ | $2 \times 10^{-1}$ $s^{-1}$ | (Howell, et al. 2000) |
| | $k_{-3}$ | $0$ $s^{-1}$ | (Ibrahim, et al. 2008b) |
| | $k_{-4}$ | $2 \times 10^{-2}$ $s^{-1}$ | (Ibrahim, et al. 2008b, Ibrahim and Henze 2014) |
| | $k_{-5}$ | $0$ $s^{-1}$ | (Ibrahim and Henze 2014) |
| | $k_{-6}$ | $1 \times 10^{-2}$ $s^{-1}$ | (Ibrahim, et al. 2008a, Ibrahim, et al. 2009) |
| | $k_{-7}$ | $1 \times 10^{-1}$ $s^{-1}$ | (Ibrahim, et al. 2008a, Ibrahim, et al. 2009) |
| | $k_{-8}$ | $8 \times 10^{-2}$ $s^{-1}$ | (Ibrahim, et al. 2008a) |
| | $k_{-9}$ | $8 \times 10^{-2}$ $s^{-1}$ | (Ibrahim, et al. 2008a) |
| | $k_{-10}$ | $1 \times 10^{-2}$ $s^{-1}$ | This study |
| | $k_{-11}$ | $1 \times 10^{-2}$ $s^{-1}$ | This study |
| | $k_{-12}$ | $1 \times 10^{-2}$ $s^{-1}$ | This study |
| | $k_{-13}$ | $1 \times 10^{-2}$ $s^{-1}$ | This study |
| **Initial amount** | | | |
| | Cdc20 | $0.22$ μM | (Fang 2002, Stegmeier, et al. 2007) |
| | O-Mad2 | $0.15$ μM | (Howell, et al. 2000, Ibrahim, et al. 2008a, Lohel, et al. 2009) |
| | Mad1:C-Mad2 | $0.05$ μM | (De Antoni, et al. 2005, Ibrahim, et al. 2008a, Lohel, et al. 2009) |
| | BubR1:Bub3 | $0.13$ μM | (Burton and Solomon 2007, Fang 2002, Ibrahim, et al. 2008a) |
| | APC | $0.09$ μM | (Stegmeier, et al. 2007) |
| | MCF2 | | (Eytan, et al. 2008), this study |
| | Other species start from zero | | |

**Table 2: *In-silico* mutation experiments for validation**

| Species | Exp. | Experimental effects | Effects in the model variants | | | | | |
|---|---|---|---|---|---|---|---|---|
| | | | MCC | BubR1 | BubR1:Cdc20 | Mad2:Cdc20 | MCF2 | All pathways |
| **Mad2** | D | Cells are unable to arrest and impaired SAC (e.g., (Dobles, et al. 2000,Fang, et al. 1998,Michel, et al. 2001,Nath, et al. 2012,Nezi, et al. 2006)) | Failed to arrest | Failed to arrest $k_{11} < 10^{-4} *k_8$ $k_8, k_{11}<10^5 M^{-1}s^{-1}$ | Failed to arrest $k_{12} < 10^{-2} *k_8$ $k_8, k_{11}<10^7 M^{-1}s^{-1}$ | Failed to arrest | Failed to arrest $k_{13} < 10^{-4} *k_8$ $k_8, k_{13}<10^5 M^{-1}s^{-1}$ | Failed to arrest all pathways $< 10^5 M^{-1}s^{-1}$ |
| **Mad2** | O | Activates the SAC and blocks mitosis and stabilizes microtubule attachment ((De Antoni, et al. 2005,He, et al. 1997,Kabeche and Compton 2012)) | Arrested | Arrested | Arrested | Arrested | Arrested | Arrested |
| **BubR1** | D | SAC dysfunction (Chan, et al. 1999,Davenport, et al. 2006,Harris, et al. 2005,Ouyang, et al. 2002) | Failed to arrest | Failed to arrest | Failed to arrest | Failed to arrest $k_{11} < 10^{-4} *k_8$ $k_8, k_{10}<10^5 M^{-1}s^{-1}$ | Failed to arrest $k_{11} < 10^{-4} *k_8$ $k_8, k_{13}<10^5 M^{-1}s^{-1}$ | Failed to arrest All pathways $< 10^5 M^{-1}s^{-1}$ |
| **BubR1** | O | Chromosomal instability (Yamamoto, et al. 2007) | Arrested | Arrested | Arrested | Arrested | Arrested | Arrested |
| **Cdc20** | D | Cells arrested in metaphase (Mondal, et al. 2006,Shirayama, et al. 1999,Zhang and Lees 2001) | Arrested | Arrested | Arrested | Arrested | Arrested | Arrested |
| **Cdc20** | O | Impairment SAC and allows cells with a depolymerized spindle or damaged DNA to leave mitosis (Hwang, et al. 1998,Mondal, et al. 2007). | Failed to arrest $k_8 < 10^5$ | Failed to arrest $k_8, k_{11}<10^6 M^{-1}s^{-1}$ | Failed to arrest $k_8, k_{12}<10^6 M^{-1}s^{-1}$ | Failed to arrest $k_8, k_{10}<10^6 M^{-1}s^{-1}$ | Failed to arrest $k_8, k_{13}<10^6 M^{-1}s^{-1}$ | Failed to arrest |

**D** refers to deletion or knockdown experiment, and **O** refers to an over-expression experiment.
Failed to arrest means very high level of APC:Cdc20 and low sequestration level of Cdc20. Arrested means very low level of APC:Cdc20 and fully sequestration of Cdc20.